\documentclass[aps,prl,twocolumn,showpacs]{revtex4}

\usepackage{graphicx}
\usepackage{amsmath}
\usepackage{amssymb}
\usepackage{bm}

\newcommand{\sinc}{{\rm sinc}}
\graphicspath{{pictPRL/}{pict/}{}}
\newcommand\pictc[5]{\begin{figure}
                       \centerline{\vspace{0mm}
                          \includegraphics[width=#1\columnwidth,height=0.7\textheight,keepaspectratio]{#3}}
                       \protect\caption{\protect\label{fig:#4} #5}\vspace{0mm}
                    \end{figure}            }
\newcommand\pict[4][1]{\pictc{#1}{!tb}{#2}{#3}{#4}}
\newcommand\rpict[1]{\ref{fig:#1}}

\newcommand\leqt[1]{\protect\label{eq:#1}}
\newcommand\reqtn[1]{\ref{eq:#1}}
\newcommand\reqt[1]{(\reqtn{#1})}
\newcounter{Fig}

\newcommand{\be}{\begin{equation}}
\newcommand{\ee}{\end{equation}}

\begin{document}

\title{Spontaneous Parametric Down-Conversion and Quantum Walks\\in Arrays of Quadratic Nonlinear Waveguides}

\author{Alexander~S.~Solntsev}
\author{Andrey~A.~Sukhorukov}
\author{Dragomir~N.~Neshev}
\author{Yuri~S.~Kivshar}
\affiliation{Nonlinear Physics Center and Center for Ultra-high bandwidth Devices for Optical Systems (CUDOS), Research School of Physics and Engineering, Australian National University, Canberra ACT 0200, Australia}

\begin{abstract}
We analyze the process of simultaneous photon pair generation and quantum walks realized by spontaneous parametric down conversion of a pump beam in a quadratic nonlinear waveguide array. We demonstrate that this flexible platform allows for creating quantum states with different spatial correlations. In particular, we predict that the output photon correlations can be switched from photon bunching to antibunching controlled entirely classically by varying the temperature of the array or the spatial profile of the pump beam.
\end{abstract}


\maketitle

Spontaneous parametric down-conversion (SPDC) is probably the most commonly used process for generation of quantum correlated photons\cite{Ribeiro:2010-87:PRsoPL} with many applications including quantum cryptography~\cite{Ekert:1992-1293:PRL} and quantum logic devices~\cite{OBrien:2003-264:NAT, Gasparoni:2004-020504:PRL}. However, the use of bulk optics for generating correlated photons as well as for the building blocks of logic gates hinders the scalability of the quantum circuitry with increasing number of components. Indeed, the successful operation of a quantum optical circuit requires that the fidelity of the quantum interference, which lies at the heart of single-photon interactions, is preserved after passing through all optical components. Integrated optical quantum circuits have been seen as the solution for on-chip scalable quantum networks with important demonstrations of multi-photon entanglement~\cite{Matthews:2009-346:NPhot}, quantum factoring algorithms~\cite{Politi:2009-1221:SCI}, and polarization entanglement~\cite{Sansoni:2010-200503:PRL}. Additionally, integrated circuits are compact, stable, and could lead in the near future to mass production of chips for quantum computation.

A particularly important platform for quantum manipulation and integrated control of non-classical light is provided by an array of coupled optical waveguides~\cite{Lederer:2008-1:PRP}. Waveguide arrays have been used to perform quantum walks of photon pairs resulting in nontrivial quantum correlations at the array output~\cite{Bromberg:2009-253904:PRL, Peruzzo:2010-1500:SCI}. Such kind of correlated walks involving quantum interference of several walkers can provide a speed-up of quantum algorithms delivering an exponential acceleration with the number of correlated walkers~\cite{Shenvi:2003-52307:PRA, Hillery:2010-62324:PRA}.
However, in all schemes to date, the correlated photon pairs were generated externally to the array by using bulk photonic elements. Such bulk elements may introduce quantum decoherence and impose stringent requirements on the losses associated with the connection of the array to the photon sources.

In this Letter, we propose and demonstrate numerically a novel scheme of quantum walks, involving {\em simultaneous} generation of correlated photon pairs through SPDC and their quantum walks inside a single photonic element~-- an array of quadratic nonlinear waveguides. This scheme avoids entirely the need for complex interfaces required in previous experiments~\cite{Peruzzo:2010-1500:SCI}, but most importantly enables novel ways for control of the spatial quantum correlations at the array output.
In particular, we show that by simply varying the phase-matching conditions for the SPDC process or the spatial profile of the pump beam it is possible to control the output quantum states incorporating photon bunching or anti-bunching. Importantly, such unusual quantum statistics are not possible when the photon pairs are created externally to the array.
Although integrated photonic couplers~\cite{Zhang:2007-10288:OE, Zhang:2010-64401:JJAP} and circuits~\cite{Saleh:2010-736:IPJ} incorporating SPDC were proposed previously, we emphasize that integrated SPDC and quantum walks in a single nonlinear array lead to additional quantum interference between probabilities to generate photon pairs in different places of the array.

Arrays of quadratic nonlinear waveguides are widely explored for manipulation of optical pulses at telecom wavelengths through cascaded generation of second harmonic~\cite{Lederer:2008-1:PRP}. Here, we consider the reverse process of SPDC and study the generation of correlated photon pairs as illustrated schematically in Fig.~\rpict{Fig1}(a). To demonstrate the flexibility in controlling photon states, we consider a phase-matched near-degenerate type I SPDC, when a pump beam generates photons of the same polarization and frequencies approximately half of the pump beam frequency. Whereas non-degenerate SPDC can also occur in the array, it can be excluded through frequency filtering at the output.
In the array, the signal and idler photon states can be associated with extended Bloch waves of the form $\exp( i k^{\perp}_{s,i} n + i \beta_{s,i} z)$~\cite{Lederer:2008-1:PRP}, where $n$ is the waveguide number, $k^{\perp}_{s,i}$ and $\beta_{s,i}$ are the normalized transverse wavenumbers and propagation constants for signal and idler waves, denoted by $s$ and $i$ subscripts, respectively. The quantum walks in an array can occur due to photon tunneling between waveguides, and we consider a common case when such tunneling occurs between neighboring waveguides and its rate can be characterized by the coupling coefficients $C_{s,i}$~\cite{Lederer:2008-1:PRP, Peruzzo:2010-1500:SCI}. We assume here that the coupling coefficients are the same for signal and idler photons within the filtered frequency range, and denote $C \equiv C_{s,i}$.
Then, the propagation constants follow a general relation for waveguide arrays~\cite{Lederer:2008-1:PRP},
\begin{equation} \leqt{beta}
    \beta_{s,i} = \beta(\omega_{s,i},k^{\perp}_{s,i})=\beta^{(0)}(\omega_{s,i}) + 2 C \cos(k^{\perp}_{s,i}).
\end{equation}
Here $\beta^{(0)}$ is a propagation constant for a single waveguide, and $\omega_{s,i}$ are frequencies of signal and idler waves. We plot the characteristic dispersion curves for the signal and idler waves in Fig.~\rpict{Fig1}(b). The dependence of propagation constant on transverse wavenumber gives rise to quantum walks of photons~\cite{Peruzzo:2010-1500:SCI}, which rate is higher for larger $C$. For a pump beam with optical frequency $\omega_p \simeq 2 \omega_{s,i}$, dispersion relation analogous to Eq.~\reqt{beta} will also apply, however the corresponding coupling coefficient $C_p$ would generally have a much smaller value compared to the signal and idler waves, $C_p \ll C$, due to the weaker mode overlap between neighboring waveguides at higher frequencies~\cite{Lederer:2008-1:PRP}. In practice, $C_p L \ll 1$, where $L$ is the array length and therefore the coupling effects can be neglected for the pump beam by setting $C_p = 0$. In this case, the input pump beam profile $A_n(n)$ remains constant inside the array.

\pict{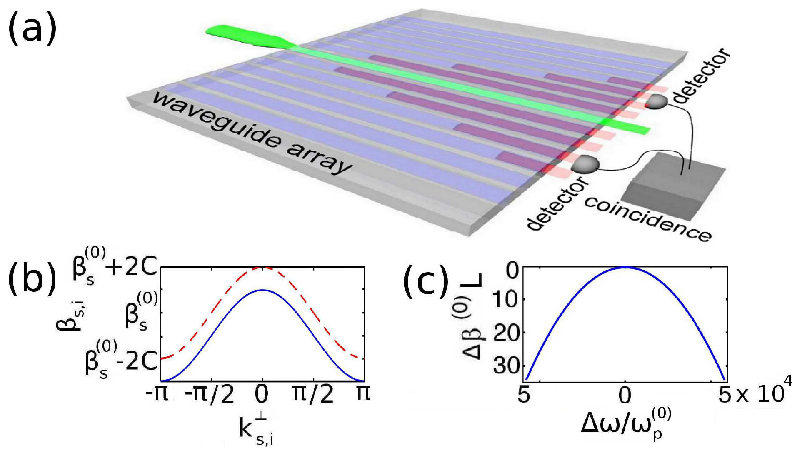}{Fig1}{(color online)
(a)~Schematic illustration of a quadratic waveguide array: the pump beam generates photon pairs that couple to the other waveguides.
(b)~Propagation constant vs. the normalized transverse wavenumber for the near-degenerate signal (red dashed line) and idler (blue solid line).
(c)~Phase mismatch vs. frequency detuning for degenerate type I oo-e SPDC in a single LiNbO$_3$ waveguide of $L=5$\,cm for extraordinary narrow-band pump at 582\,nm and ordinary signal and idler waves.
}

Now we study photon correlations at the array output by adopting the mathematical approach of Ref.~\cite{Christ:2009-33829:PRA, DiGiuseppe:2002-13801:PRA}. We describe the photon states at the array output that are associated with the extended Bloch waves, which complete set can be defined by the transverse wavenumbers from the first Brillouin zone, $-\pi \le k^\perp < \pi$. Then, the two-photon quantum state can be written as:
\begin{equation} \leqt{initial_state}
    \begin{array}{l} {\displaystyle
        | \psi \rangle = B \int_{- \pi}^{+ \pi} \! \mathrm{d}k_p^{\perp} \mathrm{d}k_s^{\perp} \mathrm{d}k_i^{\perp}
        \int_{\omega_{\rm min}}^{\omega_{\rm max}} \! \mathrm{d}\omega_s \mathrm{d}\omega_i
    }\\{\displaystyle \times
        \Omega\left(k_p^{\perp},k_s^{\perp},k_i^{\perp}\right)
        \Phi\left(k_s^{\perp},k_i^{\perp},\omega_s,\omega_i\right)
    }\\{\displaystyle \times
        \alpha\left(\omega_s+\omega_i\right) \hat{a}^{\dagger}\left(\omega_s,k_s^{\perp}\right) \hat{a}^{\dagger}\left(\omega_i,k_i^{\perp}\right) | 0 , 0 \rangle .
    } \end{array}
\end{equation}
Here $B$ is a constant, $k^{\perp}_{p}$ is a pump normalized transverse wavenumber, $n$ is a waveguide number, $(\omega_{\rm min}, \omega_{\rm max})$ is the filtered wavelength range, $\alpha$ defines pump spectrum,
$\hat{a}^{\dagger}$ are photon creation operators at the specified transverse wavenumbers and frequencies, $| 0 , 0 \rangle$ is a vacuum state.
The function $\Omega$ defines an overlap between Bloch waves,
$
        \Omega\left(k_p^{\perp},k_s^{\perp},k_i^{\perp}\right) =
        \sum_{n} A_k\left(k_p^{\perp}\right) \exp\left[\imath (k_p^{\perp} - k_s^{\perp} - \imath k_i^{\perp}) n\right],
$
where
$A_k(k_p^{\perp})$ is a $k$-space pump spectrum, found as a Fourier-transform of the input pump profile $A_n(n)$.
The function $\Phi$ in Eq.~\reqt{initial_state} defines the phase-mismatch as
$
        \Phi(k_s^{\perp},k_i^{\perp},\omega_s,\omega_i) =
        \sinc({\Delta \beta L}/{2})
        \exp( - \imath {\Delta \beta L}/{2} ),
$
where the first term defines phase-matching width, the second terms accounts for the phase accumulated by traveling photons~\cite{DiGiuseppe:2002-13801:PRA} and $\Delta \beta (\omega_{s},\omega_{i}) = \beta^{(0)}(\omega_{s}+\omega_{i})-\beta(\omega_{s},k^{\perp}_{s})-\beta(\omega_{i},k^{\perp}_{i})$, where $\beta^{(0)}(\omega_{s}+\omega_{i})$ is a propagation constant of the pump beam.

To be specific, below we consider a continuous wave or narrow-band pump with central frequency $\omega_p^{(0)}$, which spectrum can be taken as $\alpha(\omega_s+\omega_i) = \delta(\omega_p^{(0)}-\omega_s-\omega_i)$. Then, the final expression for two photon state is:

\begin{equation} \vspace{0mm} \leqt{middle_state}
        |\psi \rangle = 2 \pi B \int_{-\pi}^\pi \! \mathrm{d}k_s^{\perp} \mathrm{d}k_i^{\perp} \int_{\Delta\omega_{min}}^{\Delta\omega_{max}} \! \mathrm{d}\Delta\omega
           \left| \Psi_k (k_s^{\perp}, k_i^{\perp}, \Delta\omega) \right\rangle , \vspace{0mm}
\end{equation}
where
\begin{equation} \vspace{0mm} \leqt{Psi}
    \begin{array}{l} {\displaystyle
           \left| \Psi_k (k_s^{\perp}, k_i^{\perp}, \Delta\omega) \right\rangle  =
        A_k(k_s^{\perp} + k_i^{\perp}) \, \sinc (\Delta\beta L/2)
    }\\{\displaystyle
        \exp (-\imath \Delta\beta L/2) \,
        \hat{a}^{\dagger}(\Delta\omega,k_s^{\perp}) \,
        \hat{a}^{\dagger}(-\Delta\omega,k_i^{\perp}) \,
        | 0 , 0 \rangle . \vspace{0mm}
    } \end{array}
\end{equation}
We define the phase mismatch using Eq.~\reqt{beta}, $\Delta\beta\left(k_s^{\perp},k_i^{\perp},\Delta\omega\right)=\Delta\beta^{(0)}(\Delta\omega) - 2C\cos\left(k^{\perp}_{s}\right) - 2C\cos\left(k^{\perp}_{i}\right)$, where $\Delta\beta^{(0)}(\Delta\omega)$ is a mismatch in a single waveguide, and $\Delta\omega=\omega_s-\omega^{(0)}_p / 2=\omega^{(0)}_p / 2-\omega_i$ is a signal/idler frequency detuning from the degenerate frequency $\omega^{(0)}_p / 2$.

The phase-matching for degenerate type I SPDC can be achieved using temperature tuning in birefringent crystals with second order nonlinearity, such as LiNbO$_3$.
Waveguide arrays in LiNbO$_3$ can be routinely fabricated by Ti-indiffusion~\cite{Lederer:2008-1:PRP, Setzpfandt:2010-233905:PRL, Matuszewski:2006-254:OE}, whereas to reduce photorefraction from the strong pump beam~\cite{Matuszewski:2006-254:OE} it is possible to use MgO-doped composition~\cite{Fejer:1986-230:OL} or high sample temperatures. Characteristic dependance of phase mismatch on the frequency detuning $\Delta\beta^{(0)}(\Delta\omega)$ for the type I oo-e SPDC in near-degenerate regime is shown in Fig.~\rpict{Fig1}(c). This dependance can be approximated by the second order polynomial,
$\Delta\beta^{(0)}(\Delta\omega)=\Delta\phi^{(0)} / L - \Delta\phi^{(1)} \Delta\omega^2 / [(\omega^{(0)}_p)^2 L]$.
Here $\Delta\phi^{(0)}$ is a constant phase mismatch and $\Delta\phi^{(1)}$ corresponds to quadratic dispersion.

\pict{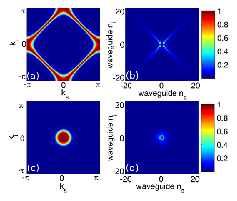}{Fig2}{(color online)
Correlations of photon pairs (a,c,e) in k space (spatial spectrum) and (b,d,f) in real space (waveguide numbers) for a
pump coupled to a single waveguide $n = 0$ with phase mismatch (a,b)~$\Delta\phi^{(0)}=0$, (c,d)~$\Delta\phi^{(0)}=40$.
}

We now calculate the second-order correlation function $\Gamma_k(k_s,k_i) = \int\mathrm{d}\Delta\omega(| \langle \Psi_k | \Psi_k \rangle | ^ 2)$, which defines correlations between photons with specific transverse wavenumbers. In order to determine correlations in real space with waveguide numbers $n_s$ and $n_i$ for signal and idler photons, we apply Fourier-transform to the two photon state $| \Psi_k(k_s^{\perp}, k_i^{\perp}, \Delta\omega) \rangle \rightarrow | \Psi_n(n_s, n_i, \Delta\omega) \rangle$. We then calculate the photon number correlation function $\Gamma_n(n_s,n_i) = \int\mathrm{d}\Delta\omega(| \langle \Psi_n| \Psi_n \rangle |^2)$, which can be measured by two detectors scanning across the waveguide array output and measuring the coincidences~\cite{Peruzzo:2010-1500:SCI} [see Fig.~\rpict{Fig1}(a)]. We take parameters for a LiNbO$_3$ waveguide array~\cite{Setzpfandt:2010-233905:PRL} $L C = 10$ such that the total structure contains about 6 coupling lengths, and consider frequency filtering $\Delta\omega_{min} \Delta\phi^{(1)} / \omega^{(0)}_p =-2$ to $\Delta\omega_{max} \Delta\phi^{(1)} / \omega^{(0)}_p =2$, which corresponds to several nm wavelength filtering for a 5\,cm long array and a pump wavelength 582\,nm. Then the required integrating time window for the detectors should be of the order of several hundred fs or longer.

We first consider the case when the pump is coupled only to the central waveguide $n=0$, which corresponds to a constant Fourier spectrum of the pump, $A_k(k_s^{\perp}+k_i^{\perp})=1$. In Fig.~\rpict{Fig2} we plot photon correlations at the array output in $k$-space and real space for different values of the phase mismatch $\Delta\phi^{(0)}$ in the range from 0 to 40 (corresponding to heating the LiNbO$_3$ array from $200$ to $202.5^{\circ}$C). For $\Delta\phi^{(0)} = 0$ we observe a square shape for the $k$-space correlations [Fig.~\rpict{Fig2}(a)]. This indicates a pronounced correlation between the generated signal and idler photons with transverse wavenumbers satisfying the relations $k_s^{\perp} \pm k_i^{\perp} \simeq \pm \pi$, which appears because these wavenumbers correspond to most efficient phase-matched interactions with $\Delta\beta = - 2C\cos\left(k^{\perp}_{s}\right) - 2C\cos\left(k^{\perp}_{i}\right) = 0$ at $\Delta\omega = 0$. We note that according to the dispersion relations~\reqt{beta}, the propagation direction can be calculated as~\cite{Lederer:2008-1:PRP} $\nu(k) = - \mathrm{d} \beta L / \mathrm{d} k = 2 C L \sin(k)$. Then, we see that at phase-matching the photons in a pair would have the same or opposite propagation directions, $\nu(k^{\perp}_{s}) \simeq \pm \nu(k^{\perp}_{i})$. Indeed, the corresponding real-space correlations shown in Fig.~\rpict{Fig2}(b) reveal that the probability of detecting signal and idler photons in either the same waveguide ($n_s=n_i$, bunching) or opposite waveguides ($n_s=-n_i$, anti-bunching) is significantly higher compared to the other probabilities. When $\Delta\phi^{(0)}$ is increased to $40$, only the generation of photons with $k_s^{\perp} \approx k_i^{\perp} \approx 0$ takes place [Fig.~\rpict{Fig2}(c)], which destroys signal and idler correlations in real space [Fig.~\rpict{Fig2}(d)].

\pict{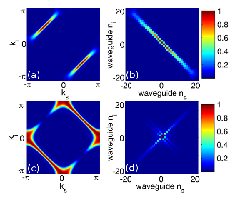}{Fig3}{(color online)
Correlations of photon pairs (a,c)~in $k$-space and (b,d)~in real space for $\Delta\phi^{(0)}=0$ and a pump coupled to waveguides $n = 0,1$ with amplitudes (a,b)~$A_n(0)=A_n(1)=1$ and (c,d)~$A_n(0)=-A_n(1)=1$.
}

The output photon statistics can be tailored by changing the pump profile and phase. When the pump beam is coupled with equal amplitudes and phases to two neighboring waveguides, $A_n(n)=1$ for $n=0,1$, then the $k$-space correlation pattern is strongly modified compared to pump input to a single waveguide [Fig.~\rpict{Fig3}(a)]. This happens because the pump spectrum is primarily concentrated in the central part of the Brillouin zone with $|k_p| \le \pi/2$, and since for phase-matched interactions $k_p \simeq k_s^{\perp} + k_i^{\perp}$, this suppresses generation of photons with wavenumbers $k_s^{\perp} + k_i^{\perp} \simeq \pm \pi$. The remaining phase-matched processes with $k_s^{\perp} - k_i^{\perp} \simeq \pm \pi$ correspond to opposite velocities of generated photons since $\nu(k^{\perp}_{s}) \simeq - \nu(k^{\perp}_{i})$, and accordingly the anti-bunching regime prevails in real space [Fig.~\rpict{Fig3}(b)]. When we introduce a $\pi$ phase difference between pump amplitudes in the input waveguides, i.e. $A_n(0)=1$ and $A_n(0)=\exp(\imath \pi) = -1$, then the situation is effectively reversed with the other phase-matched processes dominant in $k$-space, leading to bunching of signal and idler photons in real space [Fig.~\rpict{Fig3}(c,d)].

\pict[0.99]{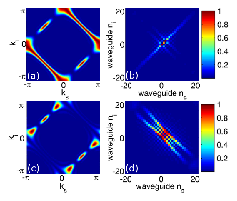}{Fig4}{(color online)
Correlations of photon pairs (a,c) in $k$-space (spatial spectrum) and (b,d) in real space (waveguide numbers) for a pump coupled to a two waveguides $n = 0,2$ (a,b) and $n = 0,3$ (c,d).
}

The output photon statistics can also be controlled by coupling the pump to spatially separated waveguides. The correlation distributions look especially interesting for the cases when waveguides $n=0,2$ or $n=0,3$ are pumped. The output $k$-space interference patterns and real space bunching or antibunching in this case depends on whether the number of non-pumped waveguides in between the pump inputs is even or odd, c.f. Fig.~\rpict{Fig4}(a-d). This is analogous to the behavior previously observed in waveguide arrays with photon pairs coupled from external source~\cite{Bromberg:2009-253904:PRL, Peruzzo:2010-1500:SCI}, however in the case of nonlinear waveguide arrays with combined SPDC and quantum walks, the correlations seem much more significantly pronounced due to the quantum interference between probabilities to create photon pairs in different places along the pumped waveguides.

In conclusion, we have studied simultaneous SPDC and quantum walks in an array of quadratic nonlinear waveguides and shown that the output correlations can be effectively controlled by changing the relative phase of the pump in the two input waveguides, as well as by altering the phase mismatch for the degenerate type I SPDC process. Such control can enable careful engineering of the output quantum state, including dynamic switching from anti- to bunching regimes. Our results provide new avenues for the development of quantum integrated circuits, combining generation of photon pairs and simultaneous transformation of the correlated photon states. It may be of interest to analyze similar systems with modulated profiles of the waveguide coupling tailored to specific requirements, or to consider type-II SPDC and generation of four-photon states~\cite{Assis:2011-3715:OE}.

We acknowledge useful discussions with G. Molina-Terriza, R. Schiek, M. Steel, and F. Setzpfandt, as well as a support from the Australian Research Council.

\end{document}